\documentstyle[psfig]{mn} 
\title[Outflows in PKS1549-79]
{Emission line outflows in PKS1549-79: the effects of the early stages of
radio source evolution?}

\author[Tadhunter et al.]
       {C. Tadhunter$^{1}$, K. Wills$^{1}$, R. Morganti$^{2}$, T. Oosterloo$^{2}$, R. Dickson$^{3}$
	\\
$^{1}$Department of Physics and Astronomy, University of Sheffield,  Sheffield, S3 7RH, UK\\ 
$^{2}$Netherlands Foundation for Research in Astronomy, Postbus 2, 7990 AA Dwingeloo, The Netherlands \\
$^{3}$Jodrell Bank Observatory, University of Manchester, Macclesfield, Cheshire, SK11 9DL. \\}

\date{}
\def\ltsim{\ifmmode\stackrel{<}{_{\sim}}\else$\stackrel{<}{_{\sim}}$\fi}
\def\gtsim{\ifmmode\stackrel{>}{_{\sim}}\else$\stackrel{>}{_{\sim}}$\fi}
\begin{document}
\maketitle
\begin{abstract}{\large We present new spectroscopic
observations of the southern radio galaxy PKS1549-79 ($z =0.152$).
Despite the flat spectrum character of the radio emission
from this source, our optical spectra
show no sign of the broad permitted lines and non-stellar continuum
characteristic of quasar nuclei and broad line radio galaxies. However, 
the high ionization forbidden lines, including
[OIII]$\lambda\lambda$5007,4959, are unusually broad for a narrow line
radio galaxy (FWHM$\sim$1350 km s$^{-1}$), and are blueshifted by 600
km s$^{-1}$ relative to the low ionization lines such as
[OII]$\lambda\lambda$3726,3729. The [OII] lines are also considerably
narrower (FWHM$\sim$650\,km s$^{-1}$) than the [OIII] lines, and have
a redshift consistent with that of the recently-detected HI 21cm absorption
line system. Whereas the 
kinematics of the [OIII] emission lines are consistent with outflow in
an inner narrow line region,  the properties of the [OII] emission lines 
suggest that they are
emitted by a more extended and quiescent gaseous component. 
We argue that, given the radio properties of the source, our line of sight is likely
to be lying close to the direction of bulk outflow of the radio
jets. In this case it is probable that
the quasar nucleus is entirely obscured at optical 
wavelengths by the material
responsible for the HI absorption line system.
The unusually broad [OIII] emission lines suggest
that the radio source is intrinsically compact. Overall, our data are
consistent the idea that PKS1549-79 is a radio source in an early stage of
evolution.}

\end{abstract}

\section{Introduction}

The near-nuclear narrow line region (NLR: $r < 5$kpc) is one of the
brightest and most readily studied components of powerful radio
galaxies. Yet, despite its potential importance for investigating gas
flows close to the central energy source, and the relationship between
optical quasar and radio jet activity (e.g. Rawlings \& Saunders
1991), we are still far from understanding its basic structure,
ionization and kinematical properties. 

One important issue concerns the dominant gas acceleration mechanism 
and the nature of the narrow line kinematics. In terms of
line widths, most powerful radio galaxies 
have [OIII] emission line widths  in the range $300
< FWHM < 600$ km s$^{-1}$ 
(Heckman et al. 1984). This is similar to
the range measured in radio-loud quasars (Brotherton 1996) and
Seyfert galaxies (Whittle 1992), and is entirely consistent with
gravitational motions in the bulges of the host galaxies (Whittle 1992).
However, a small but significant subset of radio-loud AGN show
broader [OIII] lines ($FWHM > 800$ km s$^{-1}$) which
may indicate non-gravitational motions. Interestingly,
these broader lines are invariably associated with compact
radio sources, including compact steep spectrum
(CSS) and gigahertz-peaked (GPS) sources (Gelderman and Whittle
1994). There is growing evidence that such 
compact radio sources are younger than their more extended
counterparts (e.g. Fanti et al. 1995). Therefore, 
the differences between the linewidth distributions of compact
and extended radio sources suggest that the properties
of the NLR evolve as the radio sources expand through the haloes of
the host galaxies.

Apart from the relatively crude line width measurements, more
sophisticated analyses of the line profiles reveal evidence for radial
motions in at least some classes of active galaxies. Most notably,
measurements of the line asymmetry index AI20 provide evidence for
an excess of blue asymmetries in the wings of the [OIII] lines in Seyfert galaxies
(Heckman et al. 1984, Whittle 1992). However, the evidence for systematic
radial flows in the population of radio-loud AGN is more
controversial. While Brotherton et al. (1996) found 
an excess of blue asymmetries in a sample of high redshift
radio-loud quasars, Heckman et al. (1984) found no such excess
in a sample of low
redshift radio galaxies.  

It is important to establish the direction of any radial flows. For example,
if it 
could be
shown that the gas is predominantly outflowing, then this would provide
clear evidence for non-gravitational motions associated with the
AGN. Unfortunately, 
analyses of the profiles of single emission lines do not provide a 
clear-cut answer, because the interpretation of the line 
asymmetries depends on
the (uncertain) distribution of dust in the NLR.
Given the recognition that AGN-induced outflows
may be an important feedback mechanism in formation of massive
galaxies (e.g. Silk \& Rees 1998, Fabian 1999), the issue of the
existence, or otherwise, of radial outflows has a more general
significance than the detailed phenomenology of active
galaxies.

It is clear that there is considerable uncertainty surrounding the
nature of NLR kinematics in powerful radio galaxies.  In this paper we present
spectroscopic observations of the unusual flat-spectrum radio source
PKS1549-79 which provide strong evidence for outflow in the NLR, and have a
bearing on our general understanding of the structure, kinematics and
evolution of the NLR in radio galaxies.

\section{Previous observations of PKS1549-79}

The radio source PKS1549-79 was first identified with a galaxy by
Prestage \& Peacock (1983). Subsequent spectroscopic observations by
Tadhunter et al. (1993) revealed a high ionization narrow line
spectrum, with no sign --- at least at the wavelengths covered by the
observations ($\lambda < 5500$\AA) --- of broad permitted lines that
would lead the object to be classified as a BLRG or quasar. The
redshift measured from the [OIII] emission lines in the early
observations was $z = 0.150$. However, even at the low spectroscopic
resolution ($>$20\AA\, FWHM) of the early observations it was clear
that the forbidden lines are unusually broad in this object. A further
unusual feature is that both the [OIII] lines and 5000\AA\, continuum
are polarized at the P$\sim$3\% level, indicating a significant, but
not dominant, contribution from scattered or dichroically absorbed
INLR light (di Serego Alghieri et al. 1997). The fact that a similar
degree and orientation of the polarization are measured in the lines
and continuum provides strong evidence against a non-thermal origin
for the polarized continuum.
 
PKS1549-79 is also interesting because it is one of the few known
radio galaxies which has an optical continuum that is dominated by the
light of a young stellar population (Dickson 1997, Dickson et al.
2000), with the higher order Balmer lines clearly visible in
absorption in the spectrum presented by di Serego Alighieri et
al. (1997). Another possible sign of star formation in this source is
the unusually strong far-IR emission detected by the IRAS satellite
(Roy \& Norris 1996).

Unusually for a narrow line radio galaxy (NLRG), PKS1549-79 is a
compact flat spectrum source. It is difficult to be precise about the
radio spectral index, because of the non-simultaneity of many of the
radio observations and the possibility of radio source variability,
but certainly at the higher radio frequencies the radio spectrum
appears relatively flat ($\alpha \sim 0.0$: Morganti et al. 1993). 
Moreover, the relatively
large scatter in the flux measurements at particular frequencies taken
at different epochs indicates a degree of radio variability (e.g.
Gaensler \& Hunstead 2000), although some of this variability may be
due to interstellar scintillation, given the relatively low Galactic
latitude of the source.

\begin{figure*}
\setlength{\unitlength}{1mm}
\label{fig1.fig}
\begin{picture}(10,100)
\put(0,0){\includegraphics{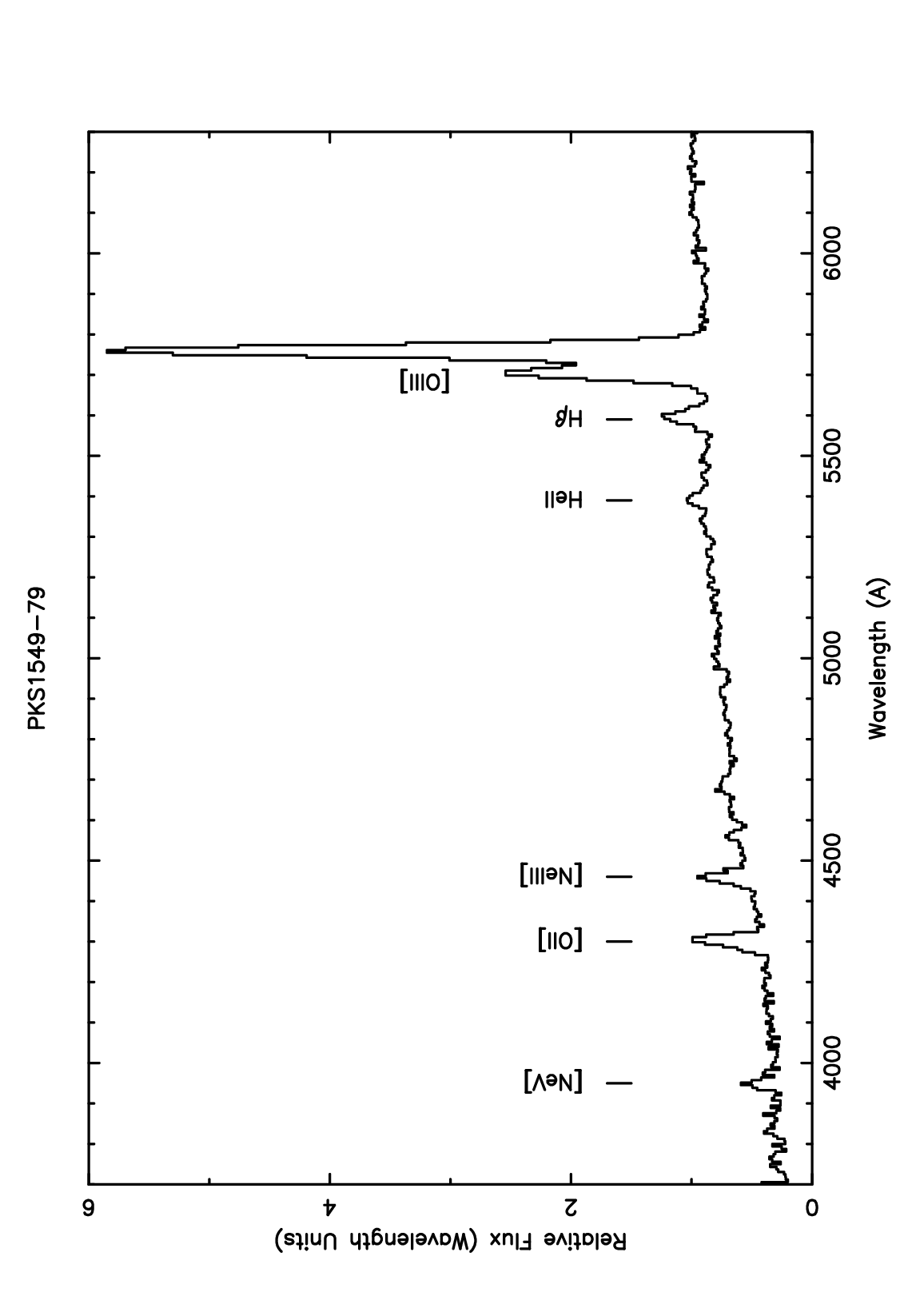}}
\end{picture}
\caption{Optical spectrum of PKS1549-79 extracted from the 1995 low resolution data.}       
\end{figure*}

In common with many flat spectrum radio sources, VLBI observations of
PKS1549-79 reveal a one-sided jet structure emanating from an
unresolved core source, with the unresolved core dominating the flux
at the higher radio frequencies (Murphy et al. 1993, King 1994). 
The jet has a distorted
structure --- bending through an angle of 60$^{\circ}$ on the NE side
of the nucleus --- and a steep radio spectrum, whereas the unresolved
radio core has a relatively flat spectrum. The total extent of the
radio source in the VLBI map is $\sim$150mas
($\sim$540pc\footnote{$H_0 =$ 50 km s$^{-1}$, $q_0 = 0.0$ assumed
throughout, resulting in a scale of 3.59 kpc arcsec$^{-1}$ for
$z = 0.152$.}), and it has been estimated that the core-jet structure
contributes 95\% of the total flux at 2.3GHz (King 1994). There is no evidence for structure on larger scales from lower
resolution maps.
Overall, the relatively small extent, one-sided
structure, flat spectrum and variability of the radio source are
similar to those observed in flat-spectrum radio sources in general.


A final interesting feature of PKS1549-79 is that significant HI 21cm
absorption has been detected against its radio core (Morganti et
al. 2001). In an attempt to gain an accurate optical emission line
redshift for comparison with the HI 21cm redshift, we took 
intermediate dispersion spectroscopic observations of 
PKS1549-79 in September 1998. These
observations are described below, along with lower spectral resolution
observations taken in 1995.

\begin{table*}
\begin{center}
\begin{tabular}{llllll}
{\bf Line} &{\bf Rest} &{\bf Relative} &{\bf Measured} &{\bf Redshift} &{\bf Line Width} \\
 &{\bf Wavelength (\AA)} &{\bf Flux} &{\bf Wavelength (\AA)} & &{\bf FWHM(km s$^{-1}$)} \\ 
\hline
{[}NeV{]} &3425.6 &43 &3942.0$\pm$1.4 &0.1506$\pm$0.0004 &1100$\pm$300 \\
{[}OII{]}(high density) &3728.8 &135 &4297.9$\pm$1.0 &0.1526$\pm$0.0002 &650$\pm$150 \\
{[}OII{]}(low density) &3728.8 &135 &4296.7$\pm$1.0 &0.1523$\pm$0.0002 &650$\pm$150 \\
{[}NeIII{]} &3868.8 &91 &4449.1$\pm$1.6 &0.1500$\pm$0.0004 &1650$\pm$220 \\
HeII  &4685.7 &40 &--- &--- &--- \\
H$\beta$ &4861.3 &100&5599.1$\pm$3.1 &0.1518$\pm$0.0004 &1700$\pm$500 \\
{[}OIII{]} &4958.9 &386 &5703.7$\pm$1.0 &0.1502$\pm$0.0002 &1420$\pm$60 \\
{[}OIII{]} &5006.9 &1156 &5758.3$\pm$1.0 &0.1501$\pm$0.0002 &1315$\pm$25 \\
{[}OI{]} &6300.3 &--- &7259.7$\pm$1.9 &0.1523$\pm$0.0003 &480$\pm$300 \\
\end{tabular}
\caption{ Emission line wavelengths, relative fluxes, redshifts and rest-frame
widths (FWHM) for PKS1549-79. 
Note that the emission line wavelengths, redshifts and line widths  
have been measured from the 1998 intermediate-resolution data, whereas
the line fluxes (presented relative to $H\beta = 100$) have been 
measured from the lower resolution
1995 data.  The relative line fluxes have an accuracy of  
approximately $\pm$10\%, and the
line
widths have been quadratically corrected for the instrumental width
(12.8\AA, FWHM). The [OII]$\lambda\lambda$3726,3728 blend has been
fitted using two Gaussians which have been constrained to have the
same width, and with the separation between the lines set by the
atomic physics. In making these fits the [OII](3726/3729) ratio has
also been constrained by making two different assumptions about the
electron density in the emitting region (high density limit and low
density limit).  
Note that the results for H$\beta$ are likely be less
accurate than for the other lines, because of the proximity of the
bright [OI]$\lambda$5577 night sky line.}
\end{center}
\end{table*}

\section{New spectroscopic observations of PKS1549-79}

The spectroscopic observations reported here were taken on the ESO
3.6m telescope with the EFOSC1 (1995) and EFOSC2 (1998)
spectroscopic/imaging instruments.  Use of the B300 grating with
EFOSC1 in 1995, and the O150 grating with EFOSC2 in 1998, resulted in
spectroscopic resolutions of 25\AA\, and 12.8\AA\, respectively. For 
the 1998
observations the
use of a slit narrower (1.5 arcseconds) than the estimated seeing disk, 
and the
alignment of the slit with the parallactic angle,  
removed any uncertainties that might result from the
different spatial distributions of the various emission lines in the
spectroscopic slit. The seeing was approximately 2 arcseconds (FWHM) for both
runs.

The data from the 1995 run were reduced following the standard steps
of bias subtraction, flat fielding, wavelength calibration,
atmospheric extinction correction and flux calibration. The reduction
for the 1998 data --- which have a generally lower S/N but a better
spectroscopic resolution --- followed the same steps, except that no
atmospheric extinction correction and flux calibration were
performed. Comparison between the results for various flux calibration
standards observed during the 1995 run reveals that the relative flux
calibration is accurate to within $\pm$5\% for the 1995 run, while
measurements of various night sky emission lines show that the
wavelength calibration is accurate to better $\pm$1\AA\, over the
entire useful wavelength range for the 1998 observations, and to
within $\pm$5\AA\, for the 1995 observations (but this assumes a
filled slit).

The observations were reduced using the Starlink FIGARO package and
analysed using the Starlink DIPSO spectroscopic analysis package.
    
\section{Results}

First we consider the narrow line spectrum based on the 1995 data,
which have a relatively high S/N and good flux calibration. Table 1
shows the relative emission line strengths derived from these data,
following subtraction of a continuum template as outlined in Dickson
(1997), while Figure 1 shows a plot of the spectrum. It is clear from
these results that the emission line ratios measured in PKS1549-79
fall within the range measured for radio galaxies in general 
(e.g. Cohen \& Osterbrock 1981), albeit
at the higher ionization end of the range.

Despite the apparently ``normal'' emission line ratios, clear
differences exist between the kinematical properties of the various
emission lines. These differences were first noticed in our lower
resolution data taken in 1990 and 1995, but were confirmed with the
higher spectral resolution data taken in 1998. Table 1 shows the
wavelengths, redshifts and rest-frame line widths measured from the
1998 data by fitting single Gaussian profiles to the stronger emission
lines.

A striking feature of the redshift measurements is that the higher
ionization lines --- [OIII]$\lambda\lambda$5007,4959,
[NeIII]$\lambda$3869, [NeV]$\lambda$3426 --- have a significantly
lower redshift ($\overline{z} =0.1501\pm0.0001$) than the low
ionization lines ­-- [OII]$\lambda$3727 and [OI]$\lambda$6300
($\overline{z}=0.1522\pm0.0002$). The redshift difference ($\Delta
z=0.0021\pm0.0002$) corresponds to a rest-frame radial velocity
difference of $\Delta v = 600\pm60$ km s$^{-1}$.  We note that the redshift
of the low ionization system is consistent with that of the HI 21cm
absorption (Morganti et al. 2000).  A further important feature is
that the high ionization lines are significantly broader ($FWHM
\sim$1350 km s$^{-1}$) than the low ionization lines ($FWHM \sim$600
km s$^{-1}$).

In order to gain further clues to the structure of the NLR in this
object, we have investigated the spatial distributions of the emission
lines relative to the continuum, by using Gaussian fits to the spatial
slices extracted from the 1995 long-slit data. The results show that
the positions of the fitted centres of [OII], [OIII] and continuum
distributions along the slit (PA270) are all consistent within the
errors ($\pm$0.1 arcseconds). However, following quadratic correction
for the seeing disk using measurements of stars along the slit, we find
that the [OII] emission is more spatially extended ($FWHM = 2.74
\pm$0.25 arcseconds or $9.8 \pm 0.9$ kpc) than the barely-resolved
[OIII] emission ($FWHM = 0.84 \pm$0.2 arcseconds or $3.0 \pm 0.72$
kpc).

How unusual are these emission line properties?  Redshift differences
between high and low ionization narrow lines appear rare, having been
noted before in only a handful of active galaxies (e.g. Koski
1978). Note, however, that the true rate of occurrence of such
redshift differences is difficult to gauge, because accurate
comparisons between the redshifts of the various lines are not always
reported when emission line spectra are presented in the
literature. What is clear is that the line widths measured in
PKS1549-79 are extreme. Very few active
galaxies at low redshifts --- fewer than 5\% of those with the accurate line width
measurements --- are known to have [OIII] lines broader than 1000 km
s$^{-1}$, and those that do are either Seyfert galaxies with powerful
linear radio sources (Whittle 1992) or, amongst radio-loud AGN,
intrinsically compact radio sources (Gelderman \& Whittle 1994).


We note in particular the remarkable similarities between PKS1549-79
and the GPS sources PKS1345+12 (4C12.30) and 3C48. Both of these
latter sources also show unusually broad [OIII] emission lines ($FWHM
= 1350$ km s$^{-1}$ in PKS1345+12 and $FWHM = 1650$ km s$^{-1}$ in
3C48: Gelderman and Whittle 1994), with the high ionization lines
significantly broader and blueshifted relative to the low ionization
lines in PKS1345+12 (Grandi 1977); both sources have unusually
luminous far-IR emission; and both sources have relatively compact
radio sources with one-sided jets. The major difference between these
two sources and PKS1549-79 is that their radio cores are weaker and
they show less evidence for variability at both low and high
frequencies.

Overall, the extreme NLR kinematics observed in PKS1549-79 have more
in common with the class of intrinsically compact radio sources ($D <
15$kpc) than with the general population of extended radio sources.

\section{Discussion}

\subsection{The structure of the NLR in PKS1549-79}

The kinematic differences between the low and high ionization lines
provide clear evidence for at least two distinct components to the NLR
in PKS1549-79: a kinematically disturbed component that is compact and
emits most of the high ionization line flux; and a more quiescent
component that has a larger spatial extent and is associated with the
HI 21cm absorption and low ionization line emission.

The most attractive explanation for the kinematic differences between
these components is that the high ionization lines are formed in a
region close to the central AGN, perhaps an inner
narrow line region (INLR) which is
undergoing systematic outflow. On the other hand, the low ionization lines are
formed in a region at greater nuclear distances which is
not disturbed kinematically by the jets or nuclear activity, and which
has a radial velocity close to systemic.

At present we cannot entirely rule out the alternative explanation:
that the [OIII] lines have a redshift close to systemic and the HI and
[OII] are associated with an infalling gas cloud or companion galaxy
which is redshifted relative to systemic. However, this explanation is
less attractive because the width the [OII] lines ($FWHM \sim$650 km
s$^{-1}$) would imply a large gravitational mass for the infalling
companion, yet existing ground-based imaging observations fail to
reveal any evidence for a massive companion galaxy along the line of
sight; the nearest object visible in ground-based images is situated
6.7 arcseconds (24kpc) to the West (Prestage \& Peacock 1983) of the
nucleus of PKS1549-79, but our spectroscopic observations show that
this is a Galactic star. 
It is also unlikely that the spatial centres of the [OII] and [OIII] emission
line distributions along PA270 would match up so accurately if the
lines were emitted by separate galactic systems.


In an attempt to gain more information about the ionization states of the two kinematic components 
using the [OII](3727)/[OIII](5007) diagnostic ratio, we have made two component Gaussian 
fits to the [OIII]$\lambda\lambda$5007,4959 and [OII]$\lambda$3727 lines, in order to 
determine the maximum 
contributions of, respectively, the broad component to [OII] and the narrow 
component to [OIII]. We find that, 
for acceptable fits to the line profiles, the 
narrow component contributes $<$25\% of the total flux of  
[OIII], while the 
broad component contributes $<$35\% of the [OII] flux. Then, using the line ratio 
information from Table 2, we find that $[OII](3727)/[OIII]5007) < 0.05$ for the broad 
component, and $[OII](3727/[OIII]5007) > 0.6$ for the narrow component.

The upper limit on [OII](3727)/[OIII](5007) for the broad 
component is unusually small for a radio galaxy. Indeed, in a recent spectroscopic survey 
of 20 intermediate redshift radio galaxies by Dickson et al. (2001), none of the objects 
showed such a small value for this line ratio. The small [OII](3727)/[OIII](5007) 
might imply an 
unusually high ionization state, a high density and/or a 
large reddening for the region emitting the broad 
lines. However, the relatively large [NeV](3426)/[NeIII](3869) and HeII(4686)/H$\beta$ 
ratios measured from the low resolution spectrum (Table 1) suggest that the
broad component has a genuinely high ionization state.
Overall, the emission line spectrum 
of the broad component appears more consistent with AGN photoionization models that 
include matter-bounded components (Binette et al. 1996), than with shock models (e.g. 
Dopita \& Sutherland 1996). 

Finally we note that the [OII](3727)/[OIII](5007) ratio deduced for the 
narrow component suggests a moderately low ionization state, but without further 
information we cannot distinguish between AGN photoionization, shock ionization or 
stellar photoionization for this component.

\begin{figure}
\label{fig2.fig}
\psfig{figure=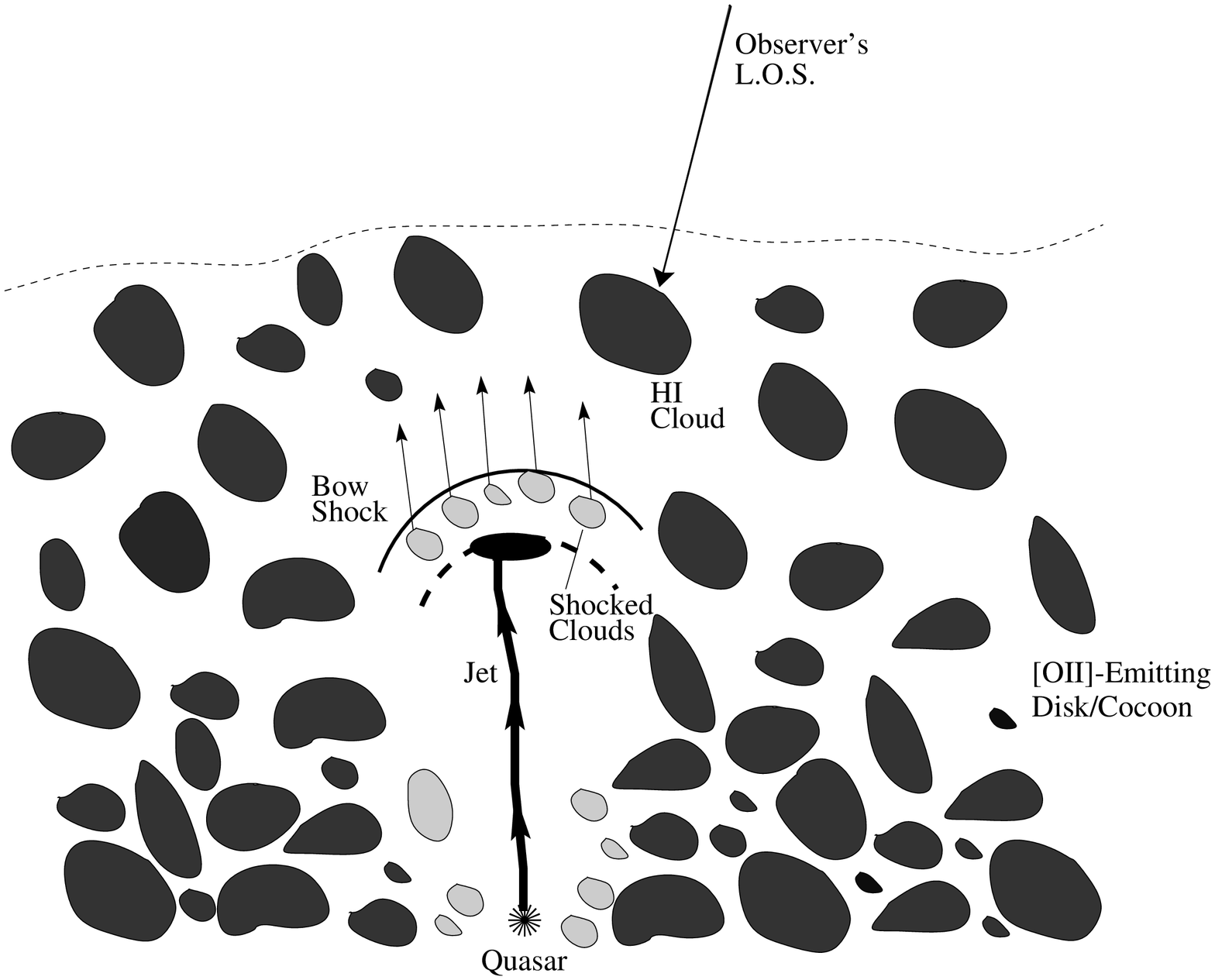,width=8.5cm,angle=270.0}
\caption{A schematic diagram showing a plausible geometric arrangement
for the various emitting components in PKS1549-79. The [OIII]-emitting
clouds are shaded light grey; most of the [OII] is emitted by material
in the extended disk/cocoon (shaded black).}
\end{figure}

\subsection{PKS1549-79 and radio source populations}

In terms of
the orientation-dependent unified schemes for powerful radio galaxies (e.g. Barthel 1989), 
we would expect PKS1549-79 to show the optical
characteristics of classes of radio-loud active galaxies which share
its radio properties, such as BL Lac objects, blazars or
quasars. However, PKS1549-79 does not fit in easily with any of these
optical classifications. Not only does it lack the non-thermal optical
continuum chracteristic of blazars and BL Lac objects, but its narrow
emission lines are much stronger than expected for a BL Lac object;
the absence of broad permitted lines lines also rules out a quasar
classification.

It is important to consider how the optical and radio propertes of
PK1549-79 can be reconciled. A possible clue is provided by the HI
21cm absorption feature, which has a similar redshift to the narrow,
low ionization emission line component. As discussed by Morganti et
al. (2001), the HI absorption characteristics are consistent with an
absorbing column density in the range $3.6\times10^{20} < N(HI) <
2.9\times10^{22}$ cm$^{-2}$, depending on the spin temperature 
($100 < T_{spin} < 8000$K). For
normal gas/dust ratio and dust extinction properties, this translates
into a visual extinction in the range $0.23 < A_v < 18$ magnitudes. At
the lower end of this range the central AGN would be lightly
extinguished, whereas at the upper end of extinction range, it would
be wholly extinguished at optical wavelengths (see Simpson et al. 1999). 
Therefore the most
plausible explanation for the optical properties of this source is
that the blazar, quasar or BL Lac nucleus is extinguished by dust
associated with the ISM that also produces the HI absorption and
narrow emission line component. In this case the broad, high
ionization emission line component may be photoionized by the hidden
nucleus, but be distributed such that it does not suffer the same
degree of extinction.

At first sight this explanation appears inconsistent with
the simplest versions of the unified schemes, which hold that we
should have a relatively unobscured view of the optical nucleus when
the jet is pointing close to the line of sight. Recently, however, the
detection of a population of relatively red flat-spectrum quasars has
led to the suggestion that dust extinction may nonetheless be
important in flat spectrum quasars (Webster et al. 1995 but see Baker
1997). Our results for PKS1549-79 support this suggestion, although
other reddening mechanisms, such as a contributions from red,
non-thermal synchrotron sources (Serjeant \& Rawlings 1996), and
contamination by the light of the stellar populations of the host
galaxies (Benn et al. 1998), are also likely to be important at some
level.

Whether or not we accept the variable dust extinction model for the
population of flat spectrum radio sources, it is clear that amongst
such sources the quasar nucleus in PKS1549-79 is unusually highly
extinguished. This high extinction may be linked to the unusual
emission line kinematics discussed in section 4 as follows. We have
argued that the extreme emission line widths measured in PKS1549-79
are consistent with it belonging to the class of intrinsically compact
radio sources. Such sources are thought to be either objects in which
the radio jets are trapped in the central regions of the host galaxies
by an unsually dense ISM (the so-called ``frustration'' model: van
Breugel 1984), or objects observed in a relatively early stage of
their evolution, before the radio jets have expanded out of the
central regions of the galaxies (the ``youth'' model: see Fanti et
al. 1995 for a discussion). In the frustration model we would
naturally expect a substantial extinction to the central AGN, even
when the jet is pointing close to our line of sight. Less obviously,
we might also expect substantial extinction along the radio jet axis
in the youth model.  This is because, when the radio jets are first
formed, the nucleus is likely to be surrounded by a cocoon or thick
disk of material left over from the events which triggered the nuclear
activity. In at least some cases the obscuring material may cover
large fraction of the sky as seen by the central source. As the radio
source evolves, any obscuring material along the radio axis is likely
to be swept aside and dissipated by jet-cloud interactions (e.g. 
Bicknell et al. 1997) or
quasar-induced winds (e.g. Balsara \& Krolik 1993) until, eventually, 
cavities are hollowed out on
either side of the nucleus. Such cavities have been detected, for
example, in the powerful extended radio source Cygnus A (Tadhunter et
al. 1999). Before this stage is reached, however, a substantial amount
of obscuring material may be present along the radio axis, but at
larger radial distances from the nucleus than the radio source.

A possible model for PKS1549-79, which shows the geometric arrangement
of the various radio and optical components, is shown schematically in
Figure 2.

The general idea that the quasar nucleus is obscured in PKS1549-79
receives indirect support from the detection of a broad Paschen
$\alpha$ emission line in infrared observations of the similar source
PKS1345+12 (Veilleux et al. 1997).  The broad Paschen $\alpha$ line in
PKS1345+12 is thought to be emitted by a quasar nucleus which is
extinguished at optical wavelengths ($A_v > 6.5$ magnitudes). Clearly,
infrared observations aimed at detecting a similar broad Paschen
$\alpha$ feature in PKS1549-79 would provide a decisive test of the
model we have proposed for this source.

\section{Conclusions and Future Work}

Most plausibly, PKS1549-79 is an intrinsically compact radio source
that has its radio jet axis pointing close to our line of sight, and
its quasar nucleus obscured from our direct view by a large amount of
obscuring material along the radio axis. While the gas responsible for
the broad, high ionization emission line component may be photoionized
by the hidden quasar nucleus, it is likely that this component is
accelerated by AGN-induced outflows. In this case,
we have caught this source in an early stage of its evolution, as
the radio jets tunnel their way through the cocoon of debris left over
from the events which triggered the activity.

A major implication of this work is that the simplest versions of the
unified schemes, in which lines of sight close to the radio jet axis
have a relatively unobscured view of the quasar nucleus, may not
always hold for young, compact radio sources in which the jets and/or
quasar winds have not yet swept aside the warm ISM in the ionization
cones on either side of the AGN. This ties in with the mounting
evidence that the quasar nuclei in the population of compact radio
sources may suffer more extinction than in extended radio sources (e.g.
Baker \& Hunstead 1995, Baker 1998).  

Clearly, detailed spectroscopic observations of larger samples of compact radio
sources have the potential to provide key information about both the early
stages of radio source evolution and the effects of the AGN activity on the
ISM of the host galaxies.

\subsection*{Acknowledgments} This work is based on observations taken using the 
European Southern Observatory 3.6m telescope, La Silla, Chile. KW and RD acknowledge
support from PPARC. We thank the anonymous referee for useful comments on an earlier
draft of this paper.

{}

\end{document}